\documentclass[aps,prc,twocolumn,groupedaddress,floatfix,letterpaper,showpacs]{revtex4}
\usepackage{amsmath,amssymb,setspace,graphicx,amsfonts,wasysym,aas_macros}
\usepackage{units}
\newcommand{\exclude}[1]{}

\begin{document}
\title{Bremsstrahlung emission from quark stars}
\author{Jean-Fran\c{c}ois Caron}
\email[]{jfcaron@phas.ubc.ca}
\author{Ariel R. Zhitnitsky}
\affiliation{Department of Physics \& Astronomy, University of British Columbia}
\date{\today}

\newcommand{\abs}[1]{\lvert#1\rvert}
\newcommand{\sq}[0]{\ensuremath{^{\textrm{2}}}}
\newcommand{\infinity}[0]{\ensuremath{\infty}}
\hyphenation{brems-strah-lung posi-tron elec-tron}


\begin{abstract}
We calculate numerically the emissivity and surface flux of electron-electron bremsstrahlung radiation from the surface of a bare quark star.  The restricted electronic phase space due to the presence of an effective photon mass results in a strong suppression.  The emissivity and surface flux are found to be substantially smaller than those found in previous work, to the point where electron-positron pair production would remain the dominant mechanism at all temperatures in the relativistic regime.  As a consequence, $e^+e^-$ pair production remains a dominant process even at low surface temperatures  $T\sim10^9$K as  originally suggested by Usov\cite{usov2}.
\end{abstract}

\maketitle

\section{Introduction}
It has been suggested long ago that the ground state of matter at large density could be the strange quark matter rather than hadronic matter as we know it~\cite{bodmer,witten84}. Strange matter is a high-density phase of QCD consisting of up, down, and strange quarks plus some electrons (to guarantee charge neutrality). It is conjectured that it is  the  ground state of matter \cite{madsenolesen,farhijaffe} (for an overview see e.g. \cite{madsenreview,alfordschmittetal}).  Strange matter, in principle,  can exist in lumps of arbitrary size ranging from the size of a nucleus to that of a star. In particular, bare quark stars could be entirely made of strange matter.  Strange matter could also make up some dark matter candidates \cite{witten84,madsenreview}(see also a recent proposal \cite{oakninzhitnitsky,zhitnitsky2006}.)  Strange matter could also exist in the cores of neutron stars whose central densities are high enough to support the high-density phase of QCD while their crusts would consist of conventional matter \cite{glendenningbook,evolution,quarknova,haensel,alcockfarhiolinto}.   

So far definitive proof of the existence of quark stars has proved elusive because many of their more measurable parameters such as spin frequency and mass overlap with neutron stars.  The most striking feature that could distinguish between quark stars and neutron stars is that quark stars have no minimum mass as mentioned above.  Neutron stars are gravitationally bound, meaning that gravity balances out the degeneracy pressure of the nucleons inside.  A consequence is that a stable neutron star must have a mass above a certain value, otherwise it would blow itself apart.  Quark stars are bound by the strong QCD force, or so called ``self-bound''.  Gravity still plays an important role in the interior details, especially for the heavier quark stars, but it is no longer essential for stability \cite{haensel}.  The result is that quark stars may have arbitrarily low masses.  Indeed ``strangelet'' dark matter candidates are effectively very small  quark stars \cite{madsenreview}.  The observation of very low mass pulsars would indicate the existence of quark stars \cite{alcockfarhiolinto,haensel}.  Unfortunately such objects would be very dim and are unlikely to be detected.

In an attempt to distinguish quark stars and neutron stars using conventional astronomy (electromagnetic radiation) in a manner that is less dependent on the details of the quark matter, radiation from the electrosphere is a natural field of interest.  Two primary radiation mechanisms have been studied so far.  These are elec\-tron-posi\-tron pair production \cite{usov2,usov,pageusov} ($e^+e^-$) and elec\-tron-elec\-tron brems\-strahlung \cite{jaikumar}.  It has been argued in \cite{jaikumar} that bremsstrahlung emission is the dominant source of radiation from the electrosphere for temperatures $10^8 \text{K}< T< 10^9\text{K}$, while $e^+e^-$ pair production is sub-dominant in this range.

The main goal of the present  work is to re-examine the radiation from the electron-electron bremsstrahlung process  using a full consideration of the phase space constraints imposed by the degenerate system.  The electrons outside the bare surface of a quark star are highly degenerate ($\mu \gg T$). Degenerate electron gasses are filled with electromagnetic waves in thermodynamic equilibrium. It results in a large plasma frequency $\omega_p$ (see Eq.\ \ref{plasmafreq}) which behaves as an effective photon mass \cite{ratkovicduttaprakash}.  This means that real photons of energy $\omega < \omega_p$ cannot propagate.  The previous examination of electron-electron bremsstrahlung by Jaikumar et al.\ only partially included this effect \cite{jaikumar}.  Namely the photon energies were restricted, but the additional restrictions on the electron energies and scattering angles were neglected. We take into account these effects and demonstrate that the bremsstrahlung emission is a sub-dominant source of radiation even for $10^8 \text{K}< T< 10^9\text{K}$ while $e^+e^-$ pair production remains a dominant process even for these low temperatures. 

The paper is organized as follows. Section \ref{quarkstars} is a short review of the physical properties of quark stars which will be essential for our analysis.  In Section \ref{bremsstrahlung} we specifically discuss the differences between our computations and the analysis presented by Jaikumat et al.\ in Ref.\ \cite{jaikumar}.  In Section \ref{approximations} we describe the approximations we have made, and how the numerical computations are done. In Section \ref{emissivitysection} we compute the emissivity while in Section \ref{fluxsection} we estimate the surface flux. Section \ref{conclusion} is our conclusion.

\section{Basic Properties of Quark Stars}\label{quarkstars}
The phase diagram of high density QCD has been very active area of research in the last few years (for a recent review see \cite{alfordschmittetal} and references therein.)  There are a number of possible QCD phases (most of them are color superconductors) which describe high density systems.  In most cases, the high density phase has a net positive electromagnetic charge \cite{alcockfarhiolinto,madsenreview,alfordschmittetal}.  This charge attracts electrons to maintain overall charge neutrality.  The electrons are only bound by electromagnetic interactions while the quarks are bound by the much stronger QCD force.  Consequently, the electron cloud that permeates the quark matter also extends past the edge of the star.  The part of the electron cloud that exists outside of the quark matter is called the electrosphere.  A canonical electrosphere for a typical quark star is around \unit[1000]{fm} thick and contains mostly degenerate ultra-relativistic electrons \cite{alcockfarhiolinto}.  Near the edge of the electrosphere, the density has decreased enough that the electrons are no longer degenerate.  At this point the electrons are also non-relativistic.  The structure of the electrosphere will play a central role in our discussion because the emission of radiation occurs exclusively from this region of the star.  The emission of electromagnetic radiation is also the primary cooling mechanism after the star has cooled sufficiently to be opaque to neutrinos.

This canonical picture of electrosphere described above has been questioned recently \cite{thickelectrosphere}.  Specifically the authors consider the presence of electromagnetically suspended quark nuggets in the electrosphere.  Such nuggets would weaken the electric field drastically, resulting in a much less dense electrosphere which extends metres away from the quark surface.  However, more careful analysis shows that these models are less likely to be realised in nature than the canonical ones \cite{thickelectrosphere2}.   Therefore, for simplicity in what follows we assume the standard picture for the electrosphere as described in the classic quark star paper by Alcock, Farhi, and Olinto \cite{alcockfarhiolinto}. 

What\-ever the model, quark matter at the dens\-ities cons\-idered is opa\-que to most electromagnetic radiation \cite{alcockfarhiolinto}.  Namely, the density is so high that radiation is strongly attenuated before it can escape.  The density profile of the quark matter is thought to be almost discontinuous at the boundary, so that there exists no ``soft quark matter'' region which could emit radiation.  The electrons in the electrosphere also attenuate radiation, but because the densities are much lower, radiation from this region is not negligible.  Detection of quark stars (if they exist) will most likely be done via radiation coming from this region.

Quark stars may have a crust of normal matter at their surface \cite{alcockfarhiolinto}.  This crust is known to have a maximum mass \cite{glendenning2}, but this is enough to completely obscure the quark surface.  In this case, radiation from quark stars would be nearly indistinguishable from that of neutron stars.  At formation, quark stars are extremely hot with temperatures on the order of $\unit[10^{11}]{K}$ \cite{hotstars}.  The high temperature allows neutrinos to escape, blowing away much of the matter that could form a crust \cite{woosley}, so most quark stars are expected to be ``bare''.  There may be exceptions in cases where the quark star is accreting matter from elsewhere \cite{alcockfarhiolinto}, but these will not be considered.  We assume in what follows that the stars are ``bare quark stars".

The sharp boundary between the quark matter and the electrosphere in bare quark stars results in an extreme electric field \cite{alcockfarhiolinto}.  At zero temperature, this electric field is strongly screened by the presence of the degenerate electrons in the electrosphere.  At finite temperature however, the electric field can be above the critical field for the Schwinger mechanism to spontaneously produce electron-positron pairs from the vacuum \cite{schwinger}.  This is the main idea behind the result by Usov \cite{usov2} where it was shown that $e^+e^-$ pair production will be the dominant process of emission for hot quark stars.  The produced positrons will later annihilate with electrons from other pairs or in the electrosphere.  The resulting spectrum is different from that seen in neutron stars, and could thus serve as a distinguishing feature \cite{usov}.  

It is important to emphasize that the resulting radiation  is not constrained to be below the Eddington limit above which radiation pressure would blow away gravitationally-bound matter.  This is because the quark stars are ``self-bound'' objects in contrast with conventional neutron stars which are bound due to gravity.  Therefore,  quark stars are the only large astrophysical objects which can radiate above the Eddington limit without shedding mass.  Indeed it has been estimated  that radiation coming from electron-positron pair production  (mentioned above) in a hot quark star can achieve luminosities well above the Eddington limit for days \cite{pageusov}.

 \exclude{ Other methods for distinguishing quark stars and neutron stars rely on gravitational waves (e.g. \cite{haenselgravity}) or cooling mechanisms that depend on the interior details (e.g. \cite{madsenidentify}).
}
\section{Bremsstrahlung in the Electrosphere}\label{bremsstrahlung}
The main goal of this section is  to  describe the  differences between the  approach we are advocating in the present work and   the results of the previous work devoted to  the same subject \cite{jaikumar}. The main  motivation for both calculations is of course the original observation that $e^+e^-$ pair production (and subsequent hard X-ray emission) will be the dominant process of emission of hot quark stars at temperatures  $8\times 10^8 \text{K}< T< 5\times 10^{10}\text{K}$  while thermal equilibrium radiation dominates at extremely high temperature $T> 5\times 10^{10}\text{K}$~\cite{usov2}.  This could serve as a distinguishing feature of quark stars.  It was expected that at lower temperatures $T<8\times 10^8 \text{K}$ the $e^+e^-$ pair production along with conventional thermodynamical equilibrium radiation will still be the dominant processes, though explicit analytical computations suffer some uncertainties in this region.
   
This result was questioned by Jaikumar et al.\ \cite{jaikumar}, who argued that at temperatures $T< 10^9\text{K}$ the radiation from electron-electron bremsstrahlung in the electrosphere exceeds that of $e^+e^-$ pair production, drastically changing the radiation spectrum.  It was argued that bremsstrahlung becomes the dominant process in this range of temperatures.  Unfortunately  the authors failed to consider all the implications of the strong electron degeneracy, which led to an overestimate of the importance of this radiation mechanism.

In what follows, we qualitatively describe two crucial elements which eventually lead to different conclusions from that presented in \cite{jaikumar}.

\subsection{Restrictions due to $\omega_p\neq 0$}\label{phasevolume}
The presence of a plasma frequency $\omega_p\neq 0$ leads to suppression of radiation.  This is because of phase volume constraints on the incoming electrons.  In order to emit a photon with energy $\omega>\omega_p$, the incoming electrons must have at least $\omega_p$ of ``excess'' energy above their chemical potential, otherwise the final electrons will not find available energy states since their energies will be below the chemical potential.   The resulting suppression is exponential with the form $\exp{(-\omega_p/T)}$.  Typically $\omega_p \gg T$ for the regions considered, so the suppression is severe.  To illustrate the emergence of the suppression, consider an integral over some restricted one-particle phase volume
\begin{equation}
\int^\infinity_0\frac{d^3p}{(2\pi)^3}\frac{2}{e^{(\epsilon-\mu)/T}+1}\Theta(\epsilon-\mu-\omega_p).
\end{equation}
Changing variables to a dimensionless excitation energy parameter $x=(\epsilon-\mu)/T$ and taking the ultra-relativistic and degenerate approximation $\abs{\vec{p}}=\epsilon=\mu$ for the polynomial $p^2$ term, we obtain
\begin{equation}
\frac{\mu^2T}{\pi^2}\int^\infinity_{\omega_p/T}\frac{dx}{e^x+1}.
\end{equation}
Since $\omega_p > T$, we have $e^x \gg 1$ for all $x$ in the integration region.  The result is
\begin{equation}
\label{supp}
\frac{\mu^2T}{\pi^2}e^{-\omega_p/T}
\end{equation}
whereas the result without $\omega_p$ is 
\begin{equation}
\label{no-supp}
\frac{\mu^2T}{\pi^2}\ln{2}.
\end{equation}
This one-particle phase volume example illustrates how the suppression comes about.  In this work the calculation is done numerically so a similar analytical expression for the suppression of bremsstrahlung when multiple particles participate cannot be given.  However, it is quite obvious that such an exponential suppression will be always present when the restriction $\omega>\omega_p$ is imposed. This generic suppression factor must be compared with the final result for the emissivity from Jaikumar et al.\ when $m^2/2\mu_e \leq T\ll \mu_e$ (see Eq.\ (40) from \cite{jaikumar}.)  They obtain for the emissivity $Q\sim \left({\cal N}(T, \mu_e)\right)^2{\cal I} (T, \mu_e)$ where ${\cal N}(T, \mu_e)\sim \mu_e^3$ while ${\cal I} (T, \mu_e)\sim T/\mu_e^2$.  Such a non-suppressed behavior from \cite{jaikumar} is  in accordance with Eq.\ (\ref{no-supp}) when no restriction is imposed. However, it should be contrasted with the correct expression given by Eq.\ (\ref{supp}) when the phase volume is constrained by $\omega>\omega_p$.

A suppression factor for our numeric calculation cannot be given in analytical form, however the phase volume constraints imposed by $\omega_p\neq 0$ can be stated simply.  Due to the strong degeneracy, only electrons near the fermi surface will participate in a given process.  This is taken into account by integrating incoming electron energies over Fermi distributions $n_F(\epsilon)$.  The outgoing electron energies must be integrated over $\tilde{n}_F(\epsilon)\equiv \left(1-n_F(\epsilon)\right)$.

The additional presence of a plasma frequency complicates things as illustrated above.  Not only are the photon energies restricted to those higher than $\omega_p$, but the electrons, already hard-pressed to interact due to the lack of available energy states, must have extra high energy in order to create the photon.  Specifically, the total energy of the incoming electrons in the lab frame must be sufficient to place the outgoing electrons in empty states and to create a photon of high enough energy.

The effect is that instead of a na\"{i}ve integration over all electron energies, the energies are constrained to a certain region in the $\epsilon_1$-$\epsilon_2$ phase space:
\begin{equation}
(\epsilon_1\epsilon_2+\Delta) < \abs{\vec{p}_1}^2\abs{\vec{p}_2}^2,\label{energycond}
\end{equation}
where $\Delta = m^2-\tfrac{1}{2}(\omega_p+2\mu+2m)^2$ is a negative quantity.

Additionally, the angle between incoming electrons is restricted because electrons which are moving in the same direction, even at high energies, cannot emit a photon of high enough energy without violating energy or momentum conservation.
\begin{equation}
1 > \cos{\theta_{12}}>\frac{-\Delta-\epsilon_1 \epsilon_2}{\abs{\vec{p}_1}\abs{\vec{p}_2}}.\label{coscond}
\end{equation}

\subsection{Soft Photon Approximation Versus Exact Bremsstrahlung}\label{comparison}
The second difference between the present analysis and \cite{jaikumar} is numerically far less important than the difference discussed above, but still deserves mention.  In the calculation by Jaikumar et al.\ of electron-electron bremsstrahlung, Low's theorem \cite{low} for emission of soft photons was used to simplify calculations and to obtain analytic results. 
In this work, the non-trivial restrictions on the phase volume of the scattering electrons (described above) force us to use numeric techniques.  Therefore, we are able to use an exact formula for the cross-section of emission of a bremsstrahlung photon \cite{alexanian} without much complication. The expression we use is larger than the one used in \cite{jaikumar}. The difference can be explained as follows. 

Low's theorem requires that the energy of the emitted photons be much smaller than any other energy scale in the system, including momentum transfer between the electrons.  This means that two relevant regions of phase space were neglected: that where emitted photon energy is near the kinematic boundary, and that where the momentum transfer between the electrons is small.  Due to these limits imposed by Low's theorem, the authors eventually arrive at the final cross-section with the a factor of $1/E^2$ in it (see Eq.\ (36) from \cite{jaikumar}.)  If the full kinematic phase space was included there should be no such factor, and the relevant formula has instead the following behavior: $d\sigma \sim r_0^2\sim 1/m^2$ (see Eq.\ (\ref{xsec}) below.)
  
  Typical electron energies are $E\approx\mu$ so our cross-section has an overall enhancement above that used by Jaikumar et al.\ by a factor $\approx m^2/\mu^2$. This effect works in the direction opposite to the effect due to the restriction on the phase volume discussed in Section \ref{phasevolume}.  Numerically the exponential suppression $e^{-\omega_p/T}$ discussed above is much stronger than a polynomial enhancement $\approx \mu^2/m^2$.  Therefore our final results for emissivity are much smaller than the results presented in \cite{jaikumar}.  To illustrate, see Figure \ref{emissivitygraph} where the emissivity is calculated with only the enhancement present and no additional electronic phase space constraints from $\omega_p$.

In this work, an ultra-relativistic cross-section for electron-electron brems\-strahlung is used.  It is applicable in arbitrary frames of reference, provided that the electrons always remain ultra-relativistic \cite{alexanian}.  This is always satisfied because of the strong degeneracy.  The only electrons that participate are those near the fermi surface which have energy $\approx 10$ MeV.  Equation (\ref{xsec}) gives the singly-differential (in the photon energy) cross-section for emission of a bremsstrahlung photon with energy $\omega$.  Each term corresponds to emission from each electron separately.  In the ultra-relativistic approximation, the cross-terms are negligible \cite{berestetskii}.
\begin{equation}
\begin{split}
\frac{d\sigma}{d\omega} = \frac{4r_0^2\alpha}{\omega}\left(1-\frac{2}{3}\frac{\epsilon_1-\omega}{\epsilon_1}+\left(\frac{\epsilon_1-\omega}{\epsilon_1}\right)^2\right)\\
\times\left(\ln{2\nu\frac{\epsilon_1-\omega}{\omega}}-\frac{1}{2}\right)+(1\leftrightarrow 2)
\end{split}\label{xsec}
\end{equation}
where
\begin{equation}
\nu = p_1\cdot p_2 \equiv \epsilon_1\epsilon_2-\vec{p_1}\cdot\vec{p_2}
\end{equation}
The cross-section above is only a slight generalization of a textbook result \cite{berestetskii}.  The calculation is done numerically because of the complexity of the full cross-section, but more importantly because of the non-trivial restrictions on the phase volume of the scattering electrons.

In the paper \cite{jaikumar}, Jaikumar et al. describe the emission spectrum of electron-electron bremsstrahlung.  For photon energies not near the plasma frequency, they use the textbook result cross-section \cite{berestetskii}.  For those energies near the plasma frequency, they use the formula resulting from applying Low's theorem.  Only in this region are the spectra different, but they are not appreciably so.  The spectrum will not be dealt with further as we show below that radiation from electron-electron bremsstrahlung is negligible.

The calculation by Jaikumar et al.\ also considered the inclusion of the Landau-\-Pomeranchuk-\-Migdal (LPM) effect \cite{lpm}.  The LPM effect encompasses multiple scattering of radiation in high-density matter.  This is most important when the wavelength of the radiation is comparable to or longer than the spacing of scattering centers.  The result is suppressed emission of low-energy photons.  We do not explicitly include the LPM effect in this work because the suppression by the phase-space constraints is strong enough to make electron-electron bremsstrahlung negligible.  Inclusion of the LPM effect would simply suppress it further.  More importantly, this suppression is expected to be numerically similar as discussed previously \cite{jaikumar}.  It could therefore be implemented if it is needed. 

\section{Approximations}\label{approximations}
In this section we briefly discuss the approximations we have implemented in our numerics. 

The cross-section used was calculated in zero-temp\-erature vacuum \cite{alexanian}.  Ideally the cross-section should be re-calculated with proper insertion of inverse fermi functions for the final electron energy integrals.  In order to avoid this tedium, we note that the cross-section itself is slowly varying.  This allows the use of an effective degeneracy factor of two  fermi functions $\tilde{n}_F(\epsilon)$, one for each electron, evaluated at typical final electron energies at the fermi surface (see section \ref{emissivitysection}).

Ideally, the final electron energy fermi functions $\tilde{n}_F(\epsilon)$ together with the energy-momentum conserving delta function would automatically suppress bremsstrahlung from electrons below the fermi surface.  In this case the outgoing electrons would have even lower energies and the final states would be mostly filled.  Because we approximate the final state degeneracy with an effective degeneracy factor, we cannot simply integrate the initial electron energies from $m_e$ to infinity.  Rather we must integrate each from $\mu+m_e$ to infinity, making sure that we satisfy the total energy condition Eq.\ (\ref{energycond}). 

The cross-section used is only valid for photon energies not close to the hard boundary of the spectrum \cite{alexanian}.  Specifically it is valid for $\omega_{max}-\omega \ll m$.  Thus those photons near the maximum energy are ignored, but the cross-section in this region is smaller than for lower energies.  The final electrons in this case also have smaller energies, so the process is further suppressed by the final state degeneracy.  The photons considered still span a much greater range than that allowed by Low's theorem.

The plasma frequency acts as an effective photon mass.  This should be reflected by recalculating the cross-section with a massive boson propagator.  This effect is a higher order correction in $\alpha$ and is expected to be small. In addition, the process involves electrons whose energies are much higher than the plasma frequency and the intermediate photon is not required to be on-shell.  This aspect of the plasma frequency is most important at small momentum transfers between the electrons, but even here we expect it to be a small effect as it leads to a correction of order $\alpha$ as explained above. 

We should note that some low energy photons with $\omega < \omega_p$ may leave the system if they are produced within $\sim \omega_p^{-1} \sim \unit[10^{-13}]{cm}$ of the outer edge of the electrosphere.  The probability for those photons with $\omega < \omega_p$ to escape is quite small as the problem is similar to tunneling in quantum mechanics when a low energy particle can escape through a high potential  barrier by means of tunneling.  However, such a process obviously includes some exponential suppression factor.  Therefore, we expect that those few photons which do escape despite the plasma frequency will not contribute much to the overall emissivity, so they are ignored.

\exclude{
The emissivity of radiation from the electron-positron pair annihilation process should also be slightly suppressed by the presence of a plasma frequency.  The spectrum of the radiation however lends itself to being unsuppressed.  The photons emitted have an average energy of $\bar{\omega}\approx\unit[0.1]{MeV}$ \cite{pageusov}.  For the plasma frequency to be relevant, it must be greater than $\bar{\omega}$.  The chemical potential profile used (see \ref{muprofile}) shows that the regions with the highest density are also the thinnest.  Thus although in some cases $\omega_p > \bar{\omega}$, the radiation length of the photons is long enough to escape the high density region into nearby low density regions.  The photons can then escape the star.  In the case of electron-electron bremsstrahlung, those photons nearest $\omega_p$ in energy could also escape by the same argument.  However the electron-electron bremsstrahlung spectrum is more slowly varying \cite{alexanian} and the bulk of the emissivity comes from the lower energy photons.  Those few photons who do escape despite the plasma frequency will not contribute much to the overall emissivity, so they are ignored.
}
\section{Emissivity}\label{emissivitysection}
The electron-electron bremsstrahlung emissivity is expressed as the energy radiated per time per volume within the electrosphere of the star.
\begin{equation}
\begin{split}
Q = \frac{1}{2}\int \frac{d^3p_1}{(2\pi)^3}\frac{d^3p_2}{(2\pi)^3} \chi(\epsilon_1, \epsilon_2,\cos{\theta_{12}})\\
2 n(\epsilon_1)2 n(\epsilon_2)(2 \tilde{n}(\mathcal{E}))^2\label{emissivity}
\end{split}
\end{equation}
where
\begin{equation}
 n(\epsilon_i) = \frac{1}{1+e^{(\epsilon_i -\mu)/T}}, ~~ i=1,2
\end{equation}
   are simply fermi functions for the initial electrons, while the factors of $2$ in front of these account for the multiple electron spin states.  The overall factor of $1/2$ accounts for the fact that only half of the photons emitted with energy $\omega>\omega_p$ can actually escape the star.  Even with high enough energy to propagate, if a photon is emitted towards the interior of the star, it will not escape.  The function
\begin{equation}
\chi(\epsilon_1, \epsilon_2,\cos{\theta_{12}}) = \int^{\omega_{max}}_{\omega_p} \omega\left(\frac{d\sigma}{d\omega}\right)d\omega
\end{equation}
contains the first moment of the cross-section.  It is integrated over photon energies from the plasma frequency \cite{ratkovicduttaprakash}
\begin{equation}
\omega_p = \sqrt{\frac{4\alpha\mu^2}{3\pi} + \frac{4\alpha\pi T^2}{9}}\label{plasmafreq},
\end{equation} 
to the limit of applicability of our cross-section as discussed above.  The effective degeneracy factor for the final electron energies is
\begin{equation}
\tilde{n}(\mathcal{E}) = 1- {n}(\mathcal{E})= \frac{1}{1+e^{(\mu-\mathcal{E})/T}}
\end{equation}
where
\begin{equation}
\mathcal{E} = \sqrt{\abs{\vec{p}_1}\abs{\vec{p}_2}(1-\cos{\theta_{12}})+\frac{m^2}{4}(2+\frac{\abs{\vec{p}_1}}{\abs{\vec{p}_2}}+\frac{\abs{\vec{p}_2}}{\abs{\vec{p}_1}})}
\end{equation}
is the centre of mass energy.  The only relevant angle is $\cos{\theta_{12}}$ which is constrained by Eq.\ (\ref{coscond}) and the energies are integrated from $\mu+m_e$ to infinity according to Eq.\ (\ref{energycond}).  

The integrations are performed numerically using Numerical Python libraries and a custom wrapper program.

\begin{figure}
\includegraphics[width=\linewidth]{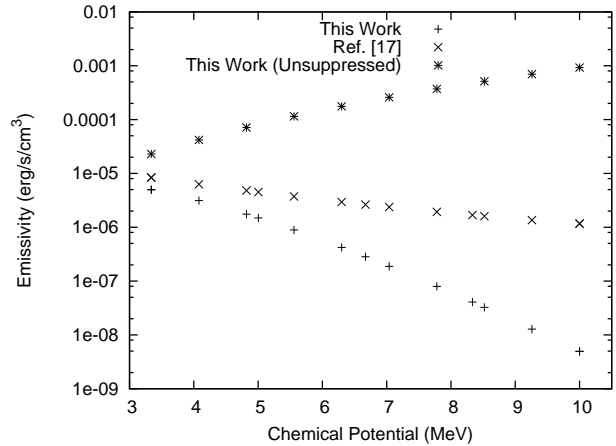}
\caption{Emissivity at $T=10^{10}$ K.  The ``Unsuppressed'' data points were calculated using the plasma frequency only as a minimum photon energy, with the full unconstrained phase volume for the electrons.\label{emissivitygraph}}
\end{figure}
The emissivity (Fig.\ \ref{emissivitygraph}) is smallest at the edge of the quark star where the chemical potential is highest.  We use a typical electron chemical potential here of \mbox{$\mu_0=10$ MeV}.  Towards the outer edge of the electrosphere, the chemical potential drops and the emissivity thus increases.  This feature is still present in the previous work, but the effects of the degeneracy were not fully included.  The result is a severely suppressed emissivity in all regions of the star for typical temperatures $T<10^{10}$K
which is due to the  phase volume suppression as qualitatively explained in Section III.

\section{Surface Flux}\label{fluxsection}
The surface flux of radiation is expressed as the energy radiated per time per surface area of the quark star.  We must integrate the emissivity along the radial direction from the bare surface of the quark star ($z=0$) to infinity.  The corresponding computations can be easily done numerically in our framework with any profile function for the chemical potential $\mu(z)$. However, in previous work \cite{jaikumar} the authors used a specific profile function to use in their analytical computations.  We duplicate their model (instead of using a known function which is a solution of mean field equations) in order to make precise comparisons with \cite{jaikumar}.
 
Previous work \cite{jaikumar} used an effective boundary of $z_0\approx \unit[1000]{fm}$. At a large enough distance the chemical potential of electrons will be so low that the electrons will no longer be ultra-relativistic.  Including the radiation from this far region is possible in principle because the density profile is known in this regime\cite{wmaphaze}.  The density here is low enough that the inclusion of radiation from this region would not affect our result for sufficiently high temperatures which is the subject of the present work, so it is neglected.

As mentioned in Section \ref{quarkstars}, modifications to the electrosphere that extend it to much larger distances than \unit[1000]{fm} \cite{thickelectrosphere} will not be discussed in this paper. In addition to the arguments presented above we should mention that these changes (even if they exist) will not affect the calculated flux at sufficiently large temperatures because the modifications occur in the non-relativistic, low density regime.  

In what follows, we assume that the temperature is constant for the entire electrosphere because energy transport is very rapid for dense degenerate matter. 
Here we use a chemical potential profile used in previous work \cite{glendenning}:
\begin{equation}
\mu(z) = \frac{\mu_0}{1+z\mu_0 \sqrt{2\alpha/3\pi}}\label{muprofile}.
\end{equation}
The surface flux is then
\begin{equation}
F = \int^{z_0}_0 Q dz.
\end{equation}
\begin{figure}
\includegraphics[width=\linewidth]{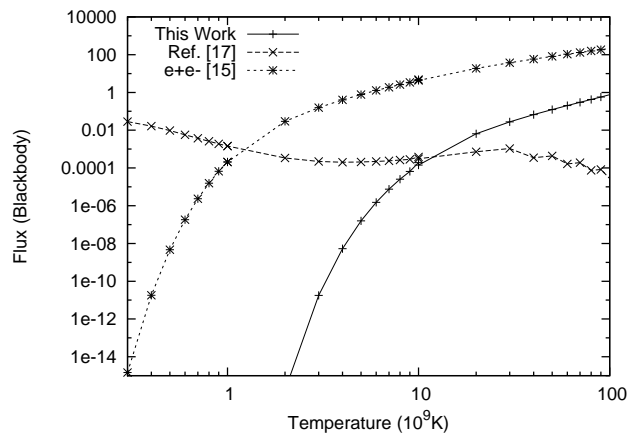}
\caption{Surface flux normalized to blackbody flux.  $\mu_0 = 10$ MeV, $z_0 = 1000$ fm.  Data points above $\unit[10^{10}]{K}$ are coarsely computed because of slow convergence.\label{Flux}}
\end{figure}
For typical temperatures, the suppression is severe to the point of making electron-electron bremsstrahlung negligible compared to electron-positron production (see Fig.\ \ref{Flux}).  At higher temperatures the suppression lessens, but only well above $10^{10}$ K.  As we mentioned in Section \ref{comparison} the use of the full kinematic phase space results in somewhat  higher emissivity than was obtained using Low's theorem in this region.  This enhancement however is not strong enough to approach the electron-positron production flux which also increases with temperature.  At these high temperatures the dominant radiation mechanism becomes thermal emission \cite{usov}, so the comparison of electron-electron bremsstrahlung and electron-positron production becomes less relevant.

At much smaller temperatures than we are considering ($T<\unit[10^8]{K}$) the region of strong degeneracy would extend all the way to the non-relativistic region.  The radiation would then come primarily from non-relativistic electrons.  In this case the analysis could be performed similarly, using a known profile function which interpolates from the relativistic to non-relativistic regime \cite{kyle2}.  The cross-section for such computations also should be modified to include relativistic and non-relativistic regimes along with the ultra-relativistic formula (Eq.\ \ref{xsec}) used in the present paper.  The results of a non-relativistic consideration would also reveal information about the radiative properties of the ``strangelet'' dark matter candidates \cite{wmaphaze,kyle2} which are effectively quark stars with a very small mass.

\section{Conclusions}\label{conclusion}
The presence of a plasma frequency in the degenerate electron gas in the electrosphere of a bare quark star has many implications on electron-electron bremsstrahlung radiation.  The restriction of the electronic phase space beyond the usual degeneracy of the electrons results in a severe suppression of the emissivity at temperatures below a few $\unit[10^{10}]{K}$. 

Full inclusion of photon energies and scattering angles beyond those permitted when using Low's theorem does not significantly increase the emissivity.  Consideration of the entire electrosphere including the non-relativistic region would further enhance the emissivity, but it would be a small contribution at sufficiently high temperatures.  The LPM effect would also suppress the emissivity of low-energy photons.  This effect has been considered previously, and therefore it is not included int our final expressions. It can be implemented if needed, but the suppression from the phase volume constraint $\omega_p \neq 0$ already reduces the electron-electron bremsstrahlung so much that the extra suppression from the LPM effect would be meaningless.

The result is that electron-electron bremsstrahlung is a negligible radiation mechanism when compared to electron-positron pair production (with subsequent $e^+e^-$ annihilation) at all temperatures considered.  This is contrary to the conclusion found in a previous consideration of the same process\cite{jaikumar}.  Electron-positron pair annihilation which results in hard X-ray emission has a very distinct spectrum compared to canonical neutron stars.  This could therefore be a powerful tool to establish the existence of quark stars and the strange matter hypothesis.

\begin{acknowledgments}
We are thankful to Volodya Usov and Kyle Lawson for correspondence and helpful discussions.  This research was supported in part by the Natural Sciences and Engineering Research Council of Canada.
\end{acknowledgments}

\bibliography{refs.bib}

\end{document}